\documentclass[aps,prd,twocolumn,superscriptaddress,nofootinbib,longbibliography]{revtex4-2}

\usepackage[T1]{fontenc}
\usepackage{newtxtext}
\usepackage{newtxmath}
\usepackage{mathrsfs}

\usepackage{amsmath}
\usepackage{mathtools}
\usepackage{bm}
\usepackage{slashed}
\usepackage{braket}

\usepackage{graphicx}
\usepackage[dvipsnames]{xcolor}
\usepackage[normalem]{ulem}

\definecolor{myred}{rgb}{0.9,0,0}
\definecolor{navy}{rgb}{0.05,0.05,0.8}
\definecolor{revisionblue}{RGB}{0,0,0}
\definecolor{deletionred}{RGB}{0,0,0}

\newcommand{\cut}[1]{}

\usepackage[colorlinks=true,citecolor=blue,linkcolor=blue,urlcolor=blue]{hyperref}
\usepackage{cleveref}

\usepackage{enumitem}
\usepackage{setspace}

\usepackage{xspace}
\usepackage{xstring}

\begin{document}

\title{Gravity-induced entanglement assisted by an additional interaction using magic angle calibration}

\author{Daisuke Miki}
\email{dmiki@caltech.edu}
\affiliation{Burke Institute for Theoretical Physics, California Institute of Technology, Pasadena, California 91125, USA}

\author{Alfred Li}
\email{ali2@caltech.edu}
\affiliation{Burke Institute for Theoretical Physics, California Institute of Technology, Pasadena, California 91125, USA}

\author{Yanbei Chen}
\email{yanbei@caltech.edu}
\affiliation{Burke Institute for Theoretical Physics, California Institute of Technology, Pasadena, California 91125, USA}

\date{\today}

\begin{abstract}
    Gravity-induced entanglement provides an operational test of whether gravitational dynamics are compatible with local operations and classical communication (LOCC), but the weakness of the gravitational interaction and the presence of force noise make direct entanglement generation extremely demanding.
    We introduce a controllable additional coherent interaction that biases two test masses close to, but within, a separability-preservation bound, so that the weak Newtonian coupling supplies only the remaining coherent contribution needed for entanglement.
    If the additional interaction alone is shown to be incapable of generating entanglement, while entanglement is observed when gravity is included, the gravitational interaction provides a non-LOCC contribution.
    Since gravity cannot be turned off, we use a magic-angle null of the projected tidal coupling to calibrate the additional interaction and force noise.
    If the additional dynamics is certified to preserve separability at the magic angle and entanglement is subsequently observed in a non-null configuration, the gravitational contribution is incompatible with an LOCC description within the Markovian model.
    The protocol trades the stringent absolute force-noise requirement of direct schemes for precision in controlling the additional coupling, its backaction, and angle-dependent systematics.
\end{abstract}

\maketitle

\textit{Introduction---}
Gravity-induced entanglement offers an operational route to testing whether gravity can mediate quantum information. Although matter-wave experiments have demonstrated the coherent response of quantum matter to an external gravitational potential~\cite{cow1975}, such observations remain compatible with a classical gravitational field. Bose et al.~\cite{bose2017} and Marletto and Vedral~\cite{marletto2017} showed that entanglement generated between initially separable systems solely by their mutual Newtonian interaction is incompatible with local operations and classical communication (LOCC). Such an observation would rule out semiclassical gravity models whose relevant bipartite matter dynamics admits an LOCC realization~\cite{kafri2013,kafri2014,diosi2017,marletto2025,marletto2025-2}. The extent to which this operational conclusion establishes quantum dynamical degrees of freedom for gravity remains under discussion~\cite{christodoulou2023,martin2023}. Related work has examined this question from several perspectives, including causality and complementarity arguments~\cite{mari2016,belenchia2018,danielson2022,mitrakos2026}, quantum-field-theoretic analyses~\cite{hidaka2022,sugiyama2023,sugiyama2024}, and S-matrix considerations~\cite{carney2022}. Recent work has also clarified the distinction between local non-entangling models and nonlocal quantum-mediated alternatives~\cite{fabiano2026}. In particular, a nonlocal entropic force model~\cite{carney2025entropic} or cross-object quantum correlations in an auxiliary monitoring channel~\cite{Monitoring2026} can supply the quantum resources required to entangle two masses. We therefore seek to distinguish gravitational dynamics that are incompatible with LOCC from LOCC descriptions by observing gravity-induced entanglement.

The central experimental obstacle of testing gravity-induced entanglement is that the gravitational coupling is weak compared with force noise in mechanical systems. Consider two systems coupled by a bilinear Newtonian interaction of strength $K_G$ and subject to force-noise processes $\mathscr F_A$ and $\mathscr F_B$, with two-sided symmetrized spectra $S_{\mathscr F_A\mathscr F_A}$, $S_{\mathscr F_B\mathscr F_B}$, and $S_{\mathscr F_A\mathscr F_B}$. Throughout this paper, we use a two-sided symmetrized spectrum $2\pi\delta(\omega-\omega^\prime)S_{\mathscr F_j\mathscr F_k}(\omega)=\braket{\{\mathscr F_j(\omega),\mathscr F_k^\dagger(\omega^\prime)\}}/2$. Within the Markovian model, a sufficient condition for separability preservation is ~\cite{kafri2013,kafri2014,diosi2017,alfred2026}
\begin{equation}
    \hbar^2 K_G^2 \le S_{\mathscr F_A\mathscr F_A}S_{\mathscr F_B\mathscr F_B}-|S_{\mathscr F_A\mathscr F_B}|^2
    \equiv N_{\mathscr F}^2.
    \label{eq:universal}
\end{equation}
Whenever this inequality holds, the same open dynamics can be reproduced by a measurement-and-feedback LOCC protocol, and hence cannot generate entanglement from any separable input state. Its violation is therefore necessary, although not generally sufficient, for entanglement generation. In the explicit two-mode protocol below, suitable state preparation and interaction time allow the dynamics to approach this threshold.

For independent Markovian white thermal noise in the high-temperature limit, the spectra are $S_{F_j^{\rm th}F_j^{\rm th}}=2M_j\gamma_jk_BT_j$ and $S_{F_A^{\rm th}F_B^{\rm th}}=0$, where $M_j$, $\gamma_j$, and $T_j$ are the mass, mechanical dissipation rate, and bath temperature of system $j=A,B$, and $k_B$ is the Boltzmann constant. For two masses separated by $d$, expanding the Newtonian potential gives $K_G=2GM_AM_B/d^3$, where $G$ is Newton's constant. For identical systems dominated by thermal noise, a necessary condition for entanglement is therefore $\hbar GM/d^3>\gamma k_BT$. We take $M/d^3\simeq20~{\rm g}/{\rm cm}^3$, which gives an axial gravity gradient comparable to that of two closely spaced coaxial gold disks~\cite{tang2025,miao2020}. The condition above then requires
$(\gamma/2\pi)T\lesssim 10^{-18}~{\rm Hz}~{\rm K}$. 
Several optomechanical protocols have been proposed for preparing suitable mechanical states and reading out the entanglement under the same necessary condition~\cite{miao2020,datta2021,miki2024,miki20242}, but the required combination of low temperature and weak mechanical dissipation remains demanding.

Various strategies have been proposed to enhance signals of gravity-induced entanglement, including the use of a massive mediator~\cite{pedernales2022}, optimized optomechanical dynamics~\cite{kaku2023,hatakeyama2026,fukuzumi2026,miki2026}, and exploiting free or unstable evolution~\cite{krisnanda2020,fujita2025,shiomatsu2025}. Complementary protocols aim to test the incompatibility of gravitational dynamics with LOCC without detecting entanglement~\cite{lami2024,mari2025}. Ref.~\cite{carney2021} further showed that an initially entangled state can boost a subsequent gravity-sensitive visibility signal. However, Ref.~\cite{alfred2026} showed that, for general bilinear interactions subject to Markovian white thermal noise, the corresponding dynamical map admits an explicit LOCC realization whenever the condition for separability preservation in Eq.~\eqref{eq:universal} is satisfied. Within these assumptions, the above strategies can increase the amount of entanglement or improve the sensitivity, but they cannot certify the non-LOCC character of gravity while Eq.~\eqref{eq:universal} remains satisfied.

In this paper, we show that a controllable additional interaction with coupling strength $K_E$ can relax the absolute force-noise requirement. The additional interaction may be electromagnetic, elastic, or otherwise engineered, provided that it couples the two masses coherently and can be calibrated independently. It is tuned so that it alone satisfies Eq.~\eqref{eq:universal}, while its coherent sum with gravity can violate the separability-preservation bound. We focus on the assisted regime $N_{\mathscr F}/\hbar \ge |K_G|$, in which gravity alone does not violate this bound. The target regime is
\begin{equation}
    \hbar^2K_E^2\le N_{\mathscr F}^2 <\hbar^2(K_G+K_E)^2.
    \label{eq:condition}
\end{equation}
The left inequality certifies that the additional interaction together with the calibrated force noise has a separability-preservation realization. The right inequality places the combined coherent interaction in a regime where suitable initial states and interaction times can generate entanglement. The additional interaction therefore biases the experiment close to the separability-preservation bound, so that the gravitational contribution determines whether the combined dynamics can become entangling. If the additional channel is independently certified in this way and entanglement is then observed in a non-null configuration, the added gravitational contribution is incompatible with an LOCC description, subject to the Markovian, additive-channel, mode-isolation, and calibration assumptions stated below.

\begin{figure}
    \centering
    \includegraphics[width=0.65\linewidth]{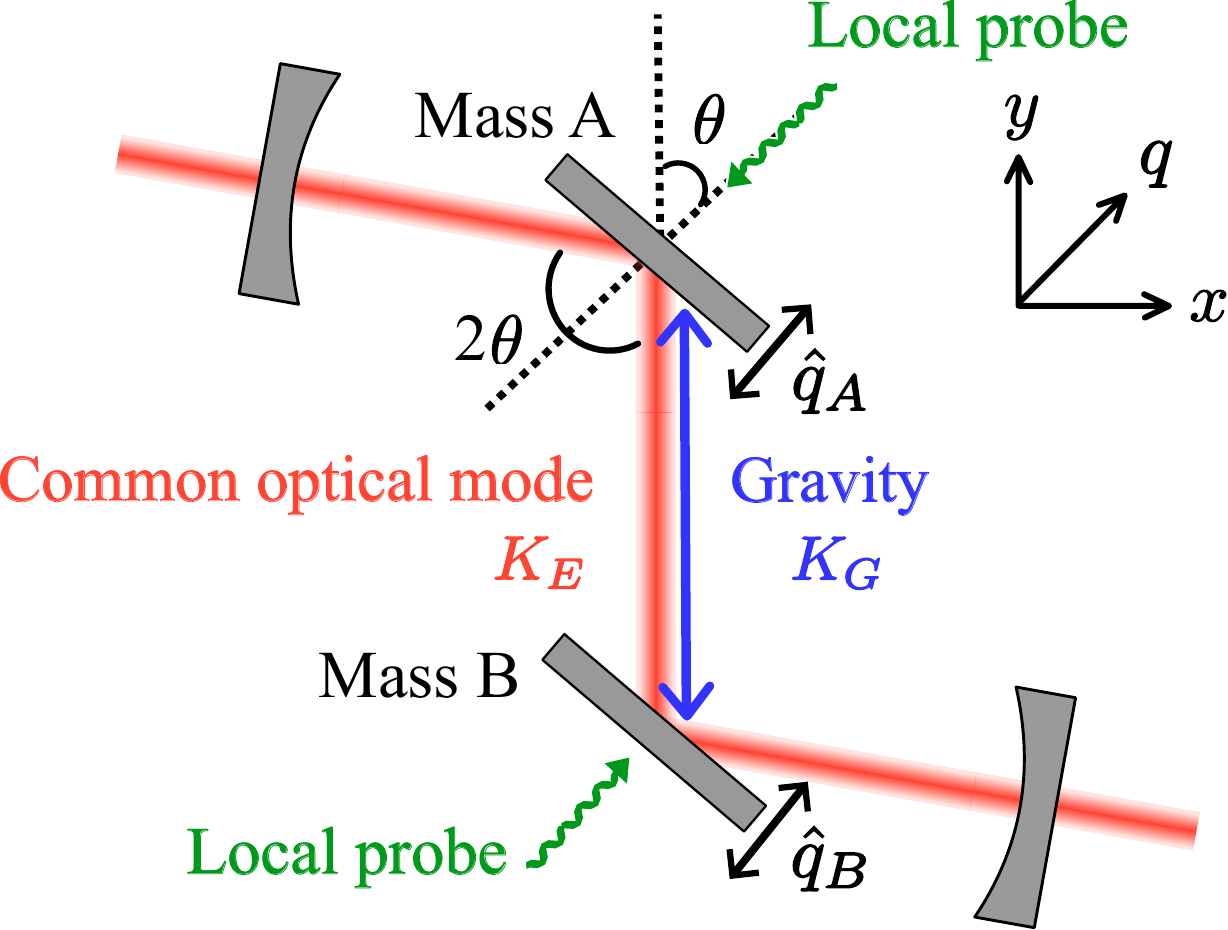}
    \caption{Schematic of the folded optomechanical setup. Two mechanical masses are separated along the $y$ axis and interact through gravity and a common optical mode. The selected mechanical coordinate $\hat q_j$ makes an angle $\theta$ with the $y$ axis. The common optical field generates the additional interaction $K_E$, and the opening angle between the two optical segments at each folding mirror is $2\theta$. The Newtonian coupling projected onto this coordinate vanishes at the magic angle, provided that coupling to the orthogonal mechanical modes is negligible. Independent local optical probes prepare and read out the two selected mechanical modes.
    }
    \label{fig:setup}
\end{figure}

Since gravity cannot be turned off, testing the left inequality requires a reference configuration in which its relevant coherent coupling vanishes. The Newtonian tidal tensor supplies such a control. In the geometry of Fig.~\ref{fig:setup}, the masses are separated along the $y$ axis, while the selected coordinate makes an angle $\theta$ with the $y$ axis. The gravitational coupling projected onto the selected modes depends on $\theta$ and vanishes at a magic angle. This does not eliminate gravity itself; gravitational coupling involving the orthogonal modes remains and must be suppressed by mode isolation. Subject to this requirement, the null configuration permits calibration of the additional interaction and force noise needed to test the left inequality in Eq.~\eqref{eq:condition}.

This interpretation assumes that if gravity were replaced by an LOCC channel, any gravity-related contribution at the magic angle either vanishes in the selected-mode projection or is independently characterized and bounded. An indistinguishable contribution that changes between the reference and operating configurations would constitute a systematic loophole. Local monitoring-feedback models~\cite{Monitoring2026} provide explicit examples in which force noise remains nonzero at the magic angle despite the vanishing projected coherent coupling.
Measuring the force auto- and cross-spectra as a function of angle could provide an independent test of the residual force noise predicted by these models.

\begin{figure*}
    \centering
    \includegraphics[width=0.45\linewidth]{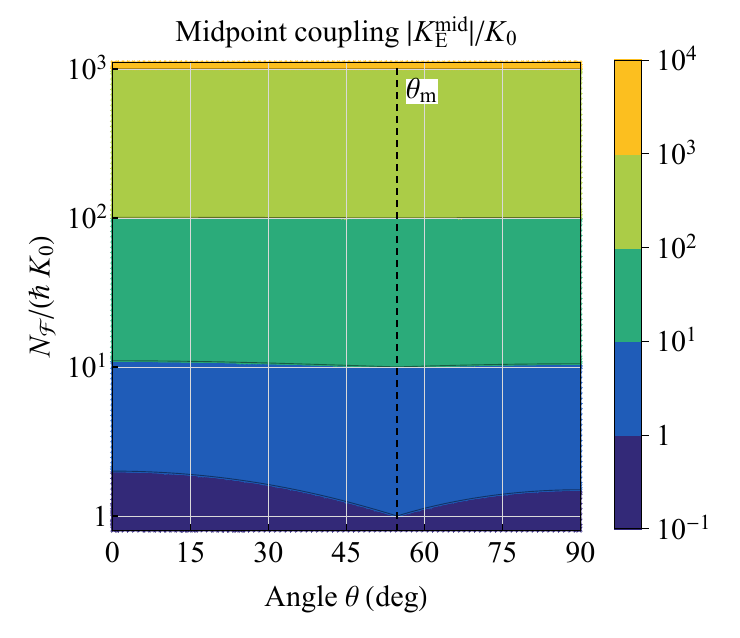}
    \includegraphics[width=0.45\linewidth]{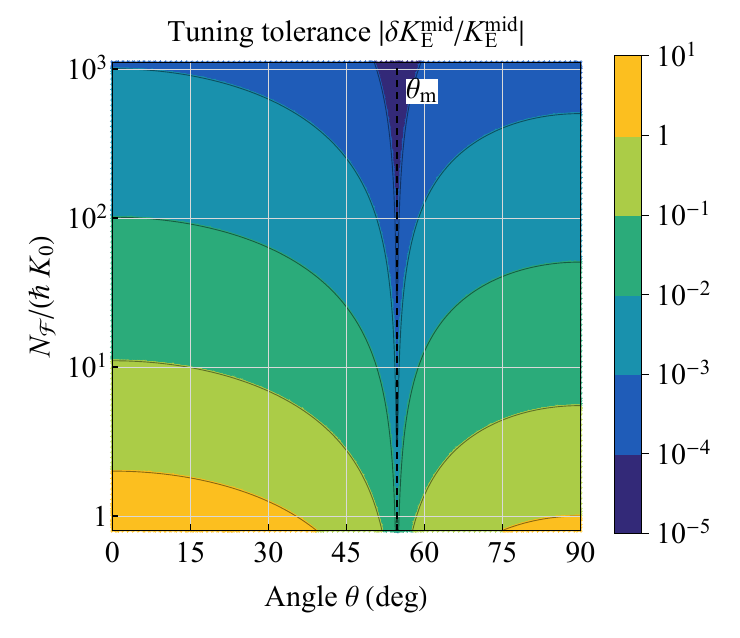}
    \caption{Assisted entanglement window and coupling-control requirement. The left panel shows the midpoint magnitude of the additional coupling $|K_E^{\rm mid}|/K_0$ as a function of $\theta$ and $N_{\mathscr F}/\hbar K_0$. The right panel shows the fractional half-width $|\delta K_E^{\rm mid}/K_E^{\rm mid}|$. The black dashed line marks the magic angle $\theta_m$. A higher permissible noise level requires a larger additional coupling but a smaller fractional calibration error, because the admissible interval in Eq.~\eqref{eq:ke required} becomes narrower relative to $|K_E^{\rm mid}|$.}
    \label{fig:additional coupling}
\end{figure*}

\textit{Dynamics---}
As shown in Fig.~\ref{fig:setup}, a common folded optical field generates the additional interaction, while independent local optical probes prepare and read out the mechanical states of test masses $A$ and $B$. Rotating the selected mechanical coordinate changes its projection onto the Newtonian tidal tensor. At the magic angle, the projected gravitational coupling vanishes, providing a reference for calibrating the
additional coupling and noise
in an entangling configuration.

As derived in Appendix~\ref{apdx:gravity}, the nonlocal part of the quadratic Newtonian potential for two-dimensional motion is
\begin{align}
    \hat H_G=K_0(2\hat y_A\hat y_B-\hat x_A\hat x_B),\quad
    K_0=\frac{GM_AM_B}{d^3},
\end{align}
where the equilibrium separation $d$ lies along the $y$ axis. The omitted local term is absorbed into the mechanical harmonic Hamiltonian as a shift of the mechanical frequency. We introduce the rotated mechanical modes as $\hat q_j=\hat x_j\sin\theta+\hat y_j\cos\theta$ and $\hat q_j^\perp=\hat x_j\cos\theta-\hat y_j\sin\theta$, where $\theta$ is measured from the $y$ axis. We assume that the selected mode $\hat q_j$ is a spectrally isolated mechanical eigenmode and that the contribution of the orthogonal mode $\hat q_j^\perp$ is negligible. The projected gravitational Hamiltonian is $\hat H_G=K_G(\theta)\hat q_A\hat q_B$ with the projected Newtonian coupling
\begin{align}
    K_G(\theta)=K_0(3\cos^2\theta-1).
\end{align}
Longitudinal motion at $\theta=0$ maximizes the magnitude of the gravitational coupling, $K_G(0)=2K_0$, whereas transverse motion gives $K_G(\pi/2)=-K_0$. The projected coupling vanishes at the magic angle
\begin{align}
    \theta_m=\arccos(1/\sqrt{3}),\quad
    K_G(\theta_m)=0
\end{align}
Thus, this magic angle $\theta_m$ is a projected null of the gravitational interaction and serves as the calibration reference.

We realize the additional interaction optomechanically. The folded optical path couples a common cavity mode $\hat a$ to both selected modes. For small displacements about the equilibrium configuration at finite $\theta$, the first-order change of the total one-way optical path length is $\delta\ell=2\cos\theta(\hat q_A-\hat q_B)$, as derived in Appendix~\ref{apdx:adiabatic}. For ideal specular reflection from plane mirrors, displacements tangent to the mirror surfaces do not contribute to this change at first order. Linearizing the cavity field about a coherent amplitude $\bar a$ gives
\begin{equation}
    \begin{split}
        \hat H_{\rm opt-int}
        &=-\hbar\mathscr G(\theta)(\hat q_A-\hat q_B)
        (\hat a+\hat a^\dagger)\\
        \mathscr G(\theta)&=2\bar a\frac{\omega_{\rm cav}}{\ell}\cos\theta
        \label{eq:linear-om}
    \end{split}
\end{equation}
where $\ell$ and $\omega_{\rm cav}$ are the cavity length and resonance frequency. Adiabatic elimination of the cavity mode gives the additional coupling and its backaction force noise $F_j^E$:
\begin{align}
    K_E(\theta)
    &=-\frac{2\hbar\Delta\,
    \mathscr G^2(\theta)}
    {\kappa^2+\Delta^2},
    \label{eq:optical-spring}\\
    S_{F_A^EF_A^E}(\theta)
    &=S_{F_B^EF_B^E}(\theta)=-S_{F_A^EF_B^E}(\theta)=\frac{4\hbar^2\kappa
    \mathscr G^2(\theta)}
    {\kappa^2+\Delta^2}S_{a_{\rm in}a_{\rm in}},
    \label{eq:optical-noise}
\end{align}
where $\Delta$ is the cavity detuning, $\kappa$ is the cavity amplitude decay rate (half width at half maximum), and $S_{a_{\rm in}a_{\rm in}}$ denotes the symmetrized noise level entering the Hermitian radiation-pressure quadrature; $S_{a_{\rm in}a_{\rm in}}=1/2$ for vacuum in our normalization. For vacuum input, we have $S_{F_A^EF_A^E}/\hbar |K_E|=\kappa/|\Delta|$. Thus, the dispersive regime $|\Delta|\gg\kappa$ suppresses backaction relative to the coherent additional coupling, at the cost of increased optical power for fixed $K_E$.

In the interaction picture of the dressed local Hamiltonian, the effective selected-mode master equation is
\begin{align}
    \dot\rho=&\frac{1}{i\hbar}
    [(K_G(\theta)+K_E(\theta))\hat q_A\hat q_B,\rho]\nonumber\\
    \qquad&
    -\sum_{j,k=A,B}\frac{S_{\mathscr F_j\mathscr F_k}(\theta)}{2\hbar^2}
    [\hat q_j,[\hat q_k,\rho]],
    \label{eq:master-equation}
\end{align}
where $S_{\mathscr F_j\mathscr F_k}(\theta)$ is the spectral density of the total force noises acting on  systems $j$ and $k$. For the sum of independent Markovian white thermal noise and optical backaction, $\mathscr F_j=F_j^{\rm th}+F_j^{E}$, the effective noise amplitude is
\begin{align}
    N_{\mathscr F}^2=S_{F_A^{\rm th}F_A^{\rm th}}S_{F_B^{\rm th}F_B^{\rm th}}+(S_{F_A^{\rm th}F_A^{\rm th}}+S_{F_B^{\rm th}F_B^{\rm th}})S_{F_A^EF_A^E}.
\end{align}

\textit{Requirements for calibration---}
At the magic angle, the projected gravitational coupling vanishes, $K_G(\theta_m)=0$, providing a reference configuration for the additional interaction. However, the additional coupling $K_E$ and the effective noise amplitude $N_{\mathscr F}$ generally depend on $\theta$. Establishing the left inequality in Eq.~\eqref{eq:condition} at $\theta_m$ therefore does not by itself imply that the additional channel preserves separability at other angles. A reference measurement at $\theta_m$ can be transferred to another angle only after this instrumental dependence is compensated or calibrated.

Eqs.~\eqref{eq:optical-spring} and \eqref{eq:optical-noise} show that, for fixed detuning, cavity bandwidth, and input noise spectrum, both $K_E(\theta)$ and $S_{F_j^EF_k^E}(\theta)$ are proportional to $\mathscr G^2(\theta)$.
We adjust the drive to keep this quantity constant during an angular scan. The same calibration must match the additional interaction and noise between the operating configuration and the magic-angle reference. Compensation becomes increasingly demanding
as $\theta$ approaches $\pi/2$ and
fails when the optical projection vanishes. Different values of $\theta$ correspond to separately aligned equilibrium configurations. In the geometry of Fig.~\ref{fig:setup}, the mirror normals are reoriented together with the suspensions, while the equilibrium separation $d$ is held fixed. The common cavity and the local preparation and readout probes are realigned accordingly. Changes in mechanical frequency and quality factor must be included in the calibration. If the thermal force spectra are also unchanged, $N_{\mathscr F}$ is independent of $\theta$.

We choose the additional coupling to satisfy $\hbar^2K_E^2\le N_{\mathscr F}^2$. At the reference angle, this condition guarantees that the additional channel preserves separability. At an operating angle where gravity adds constructively to the additional coupling, the combined interaction can satisfy $N_{\mathscr F}^2<\hbar^2(K_G(\theta)+K_E)^2$. For $K_E$ and $K_G(\theta)$ of the same sign, Eq.~\eqref{eq:condition} becomes
\begin{align}
    \frac{N_{\mathscr F}}{\hbar}-|K_G(\theta)|<|K_E|\le \frac{N_{\mathscr F}}{\hbar}.
    \label{eq:ke required}
\end{align}
The midpoint of this interval is $K_E^{\rm mid}={\rm sgn}[K_G(\theta)](N_{\mathscr F}/\hbar-|K_G(\theta)|/2)$ where ${\rm sgn}$ denotes the sign function. The left panel of Fig.~\ref{fig:additional coupling} shows $|K_E^{\rm mid}|/K_0$ as a function of angle and noise level. When $N_{\mathscr F}/\hbar\gg|K_G|$, the required magnitude of the additional coupling is approximately $N_{\mathscr F}/\hbar$. Stronger assistance therefore permits a larger noise level.

However, there is a trade-off between the allowable noise level and the required precision of the additional interaction. For a fixed, independently calibrated $N_{\mathscr F}$, define the magnitude half-width $\delta K_E^{\rm mid}=N_{\mathscr F}/\hbar-|K_E^{\rm mid}|=|K_G(\theta)|/2$. When $N_{\mathscr F}/\hbar\gg |K_G(\theta)|$, the corresponding relative tolerance approaches $|\delta K_E^{\rm mid}/K_E^{\rm mid}|\simeq\hbar|K_G(\theta)|/2N_{\mathscr F}$, as plotted in the right panel of Fig.~\ref{fig:additional coupling}. Uncertainty in the noise level must also be included in this error budget. The tolerance becomes more stringent near the magic angle because the gravitational contribution, and hence the width of the assisted window, decreases. The reference angle is useful for calibrating the additional coupling, whereas an operating angle away from it provides a finite margin for entanglement generation. 
Fig.~\ref{fig:additional coupling} also shows that $\theta=0$ gives the widest admissible interval of $|K_E|$ and permits the largest force noise for a given fractional coupling tolerance.

To quantify the relaxation of the conventional thermal-decoherence requirement, we define the dimensionless factor 
\begin{align}
    R_{\rm th} = \gamma T/\gamma_0T_0
\end{align}
where $(\gamma_0/2\pi)T_0 = 10^{-18}~{\rm Hz\,K}$ is the representative benchmark for unassisted gravity-induced entanglement.
At $\theta=0$, the midpoint is $K_E^{\rm mid} = N_{\mathscr F}/\hbar-K_0$ and its magnitude half-width is $K_0$. For $M/d^3=20~{\rm g}/{\rm cm}^3$ and thermal noise dominating optical backaction, $R_{\rm th}=10$ gives a fractional half-width $|\delta K_E^{\rm mid}/K_E^{\rm mid}|$ of approximately $9\times10^{-2}$, while $R_{\rm th}=10^3$ gives $8\times10^{-4}$.
The gain in noise tolerance is therefore exchanged for coupling stability and calibration accuracy.

\begin{figure}
    \centering
    \includegraphics[width=0.95\linewidth]{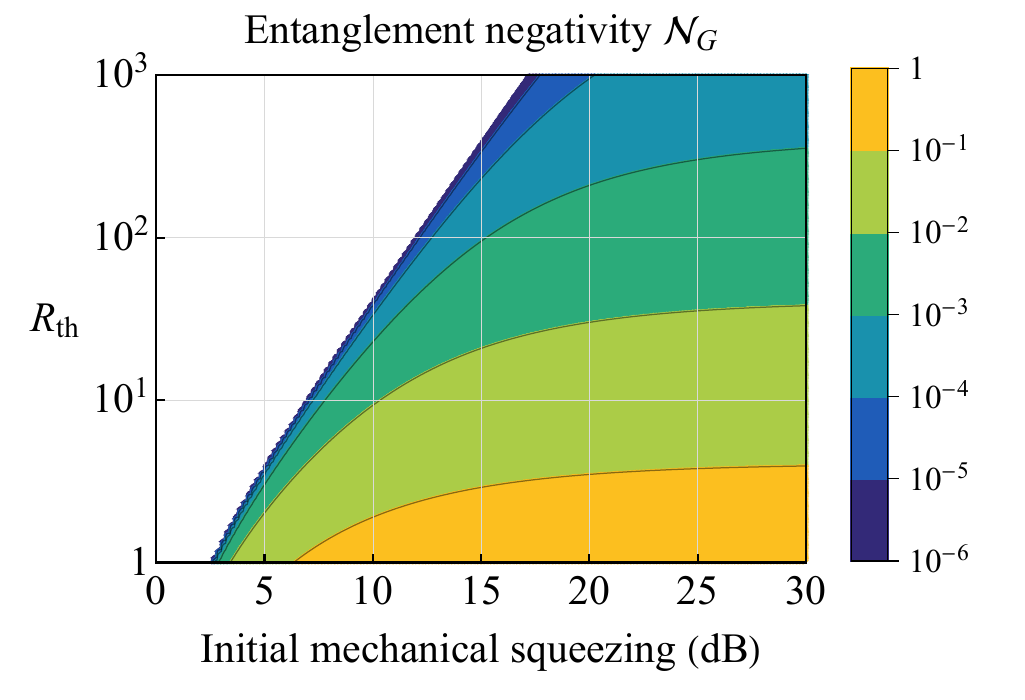}
    \caption{
    Entanglement negativity $\mathscr N_G$ read out by local optical probe pulses. The mechanical interaction duration $t_G$ is chosen to maximize the negativity at every point, following Appendix~\ref{apdx:angular dependence}. The axes show initial mechanical squeezing and $R_{\rm th}=\gamma T/\gamma_0T_0$ relative to the unassisted benchmark $\gamma_0T_0$. We assume pure initial mechanical squeezed states, $\theta = 0$ so that $K_G = 2K_0$, $K_E = K_E^{\rm mid}$, $M/d^3 = 20~{\rm g}/{\rm cm}^3$, and thermal noise dominating optical backaction. 
    The white region denotes zero entanglement $\mathscr N_G = 0$.
    Other angles are discussed in Appendix~\ref{apdx:angular dependence}.}
    \label{fig:en}
\end{figure}

\textit{Detection of entanglement---}
We separate the protocol into state preparation, evolution under gravity and the additional interaction for a duration $t_G$, and local readout. The two mechanical modes are initially prepared independently, for example using pulsed optical protocols~\cite{vanner2011,hofer2011,bassam2019} or by conditioning on continuous measurement~\cite{meng2020,matsumoto2020}. During the interaction stage, the preparation apparatus is switched off, and the masses evolve subject to thermal noise and the backaction of the additional optical field. For readout, we assume local transfer of the final mechanical states onto outgoing optical modes, preserving their entanglement. Appendix~\ref{apdx:pulsed protocol} describes a pulsed realization and the conditions for complete state transfer, following Ref.~\cite{miki2026}.

We consider identical mechanical systems and use the rotating-wave approximation (RWA).
The Hamiltonian in the interaction picture is
\begin{align}
    \hat H^I&=
    \hbar g_\theta(\hat b_A\hat b_B^\dagger+\hat b_A^\dagger\hat b_B),
\end{align}
where $g_\theta = (K_G(\theta)+K_E)/2M\omega_m$ is the coherent coupling and $\hat b_j = \sqrt{M\omega_m/2\hbar}\hat q_j+i\hat p_j/\sqrt{2M\hbar \omega_m}$ is the phonon annihilation operator. The RWA is taken in the weak-coupling regime $|g_\theta|\ll \omega_m$, with interaction times satisfying $\omega_mt_G\gg1$. The equations of motion With Markovian white force noise are
\begin{align}
    \begin{split}
        \dot{\hat b}_A&=
        -ig_\theta \hat b_B-\frac{\gamma}{2}\hat b_A+\mathscr F_{A,\rm nor}\\
        \dot{\hat b}_B&=
        -ig_\theta \hat b_A-\frac{\gamma}{2}\hat b_B+\mathscr F_{B,\rm nor},
    \end{split}
\end{align}
where $\mathscr F_{j,\rm nor}$ is normalized force noise and its correlations are $S_{\mathscr F_{A,\rm nor}\mathscr F_{A,\rm nor}}=S_{\mathscr F_{B,\rm nor}\mathscr F_{B,\rm nor}}\simeq\gamma N_{\rm th}+S_{F_A^EF_A^E}/2M\hbar\omega_m$ and $S_{\mathscr F_{A,\rm nor}\mathscr F_{B,\rm nor}}=S_{F_A^EF_B^E}/2M\hbar\omega_m$, with thermal occupation $N_{\rm th}=k_BT/\hbar\omega_m$ in the high-temperature limit.

Introducing the quadrature vector $\hat{\bm r}=(\hat b_{1A},\hat b_{2A},\hat b_{1B},\hat b_{2B})^{\rm T}$, with $\hat b_{1j}=\hat b_j+\hat b_j^\dagger$ and $\hat b_{2j}=(\hat b_j-\hat b_j^\dagger)/i$, we define the mechanical covariance matrix as $\bm V_b=\langle\{\hat{\bm r}-\langle\hat{\bm r}\rangle,(\hat{\bm r}-\langle\hat{\bm r}\rangle)^{\rm T}\}\rangle/2$. For a two-mode Gaussian state, the entanglement negativity is $\mathscr N_G={\rm max}\{(1/\nu_--1)/2,0\}$, where $\nu_-$ is the smallest symplectic eigenvalue after partial transposition. The state is entangled if and only if $\nu_-<1$, equivalently $\mathscr N_G>0$. The explicit form is shown in Appendix~\ref{apdx:pulsed protocol}. 

For a quantitative example within the RWA, we assume an initial product of two pure mechanical squeezed states, $\bm V_{b,\rm in}={\rm diag}\{e^{2\zeta},e^{-2\zeta},e^{2\zeta},e^{-2\zeta}\}$, where $\zeta\ge0$ and the squeezing in decibels is $20\zeta/\ln10$.
After an interaction time $t_G$, the smallest symplectic eigenvalue after partial transposition is
\begin{align}
    \nu_-\simeq&
    \sqrt{(\cosh2\zeta+2\gamma t_GN_{\rm th})^2-\sinh^22\zeta\cos^2(2g_\theta t_G)}\nonumber\\
    &-|\sinh2\zeta\sin(2g_\theta t_G),
    \label{eq:symplectic eigenvalue}
\end{align}
where, in addition to the rotating-wave approximation, we assume $\gamma t_G\ll1$ and thermal noise that dominates the optical backaction. This expression agrees with Ref.~\cite{miki2026} after replacing the interaction rate $g_G$ with $g_\theta$.

Figure~\ref{fig:en} shows the entanglement negativity as a function of initial mechanical squeezing and the thermal-noise ratio $R_{\rm th}$.
We take the longitudinal operating point $\theta=0$, with $K_G(\theta)=2K_0$, $K_E=K_E^{\rm mid}$, and $M/d^3=20\,{\rm g}/{\rm cm}^3$. For these parameters, entanglement generation without assistance requires $N_{\mathscr F}/\hbar <|K_G|$, corresponding to $R_{\rm th}\lesssim 1$.
We choose $t_G$ to maximize the negativity at every point, as discussed in Appendix~\ref{apdx:angular dependence}. 
As the initial mechanical squeezing increases, both the generated entanglement and the tolerable thermal noise increase. In the infinite-squeezing limit, the two-mode protocol approaches the separability boundary~\cite{miki2026}. For the moderate assisted regime $R_{\rm th}=10$ discussed above,
approximately $10\,{\rm dB}$ of initial mechanical squeezing is already sufficient to generate entanglement. Stronger assistance can tolerate substantially more thermal noise; for example, the protocol reaches $R_{\rm th}=10^3$
with approximately $20~{\rm dB}$ of initial mechanical squeezing.

The combined thermal noise, squeezing, and calibration requirements remain experimentally challenging. Although cryogenic ultrahigh-vacuum environments have been demonstrated in particle-trap systems~\cite{gabrielse2002,baglin2024}, representative low-dissipation mechanical resonators remain above the combined bath-temperature and intrinsic-dissipation scale required here~\cite{catano2020,hofer2023,agafonova2026}. Although the protocol still remains experimental difficulty, it trades the thermal-noise requirement for tighter control of the additional coupling and its backaction, resulting in improvements of several orders of magnitude.

\textit{Conclusion---}
We have proposed an assisted gravity-induced entanglement protocol in which a controllable coherent interaction biases two test masses close to a separability-preservation bound, allowing much weaker Newtonian coupling to determine whether the combined dynamics can generate entanglement. A magic-angle null of the projected Newtonian tidal coupling provides a reference for calibrating the additional interaction and force noise. Within the Markovian model, if the additional channel is certified to preserve separability at the reference configuration, its coupling and noise are matched at the operating angle, and entanglement is observed with nonzero projected gravitational coupling, the gravitational contribution is incompatible with an LOCC description.
The method changes the experimental challenge: increased force-noise tolerance is purchased with tighter control of the additional coupling, its backaction, and angle-dependent systematics. The inference also requires any gravity-related LOCC contribution at the magic angle either to vanish in the selected-mode projection or to be independently bounded. This resource trade and the tensorial null provide a complementary route to testing non-LOCC gravitational dynamics.

\textit{Acknowledgments---}
D.M.\ is supported by the Japan Society for the Promotion of Science (JSPS) Overseas Research Fellowships and the Keck Foundation. Y.C.\ and A.L.\ are supported by the Simons Foundation (Award No.~568762).

\bibliographystyle{apsrev4-2}
\bibliography{reference}

\clearpage

\appendix

\section{Newtonian expansion}
\label{apdx:gravity}

The quantum Newtonian potential is given by
\begin{align}
    \hat H_G=-\frac{GM_AM_B}{|\bm d-\hat{\bm r}_A+\hat{\bm r}_B|},
\end{align}
where $\bm d=(0,d)$ and $\hat{\bm r}_j=(\hat x_j,\hat y_j)^{\rm T}$.
Using $\delta\bm r=\bm r_A-\bm r_B$, we can expand the denominator of the potential as
\begin{equation}
    \frac{1}{|\bm d-\delta\bm r|}
    =
    \frac{1}{d}
    +\frac{\bm n\cdot\delta\bm r}{d^2}
    +\frac{3(\bm n\cdot\delta\bm r)^2-\delta\bm r^2}
    {2d^3}
    +O(\delta\bm r^3/d^4).
    \label{eq:newton-expansion}
\end{equation}
The zeroth-order term is an irrelevant constant. The first-order term produces a static force and is removed by redefining the equilibrium positions of the mechanical modes. Retaining the quadratic contribution gives
\begin{align}
    \hat H_G
    &=\frac{GM_AM_B}{d^3}\left[2\hat y_A\hat y_B-\hat x_A\hat x_B\right]+\hat H_G^{\rm local},
\end{align}
where
\begin{align}
    \hat H_G^{\rm local}
    &=\frac{GM_AM_B}{2d^3}\left(\hat x_A^2+\hat x_B^2-2\hat y_A^2-2\hat y_B^2\right).
\end{align}
The local contribution $\hat H_G^{\rm local}$ renormalizes the mechanical restoring potentials. It cannot generate bipartite entanglement, but it is included in the dressed mechanical frequencies. For two identical oscillators with equal local frequency shifts, this contribution can be absorbed into a common dressed frequency. Within the rotating-wave approximation, the corresponding common local evolution is a product of local phase-space rotations and does not change the entanglement-generation condition.

We now introduce a rotated mechanical basis as $\hat q_j=\hat x_j\sin\theta+\hat y_j\cos\theta$ and $\hat q_j^{\perp}=\hat x_j\cos\theta-\hat y_j\sin\theta$. In this basis, the gravitational Hamiltonian becomes
\begin{align}
    \hat H_G&=
    \frac{GM_AM_B}{d^3}\biggl[(3\cos^2\theta-1)\hat q_A\hat q_B\nonumber\\
    &\qquad
    -3\sin\theta\cos\theta(\hat q_A\hat q_B^\perp+\hat q_A^\perp\hat q_B)\nonumber\\
    &\qquad
    +(2-3\cos^2\theta)\hat q_A^\perp\hat q_B^\perp\biggr]+\hat H_G^{\rm local}
\end{align}
We assume that $\hat q_j$ denotes a spectrally isolated mechanical eigenmode of each resonator. The orthogonal translational, rotational, and internal modes are assumed to be sufficiently far detuned or sufficiently strongly confined that their contributions to the selected mode are negligible. Such isolation may be supplied by the mechanical suspension and, if necessary, strengthened by additional local optical trapping fields.

Under this single-mode projection, the gravitational Hamiltonian reduces to
\begin{align}
    \hat H_G=\frac{GM_AM_B}{d^3}(3\cos^2\theta-1)\hat q_A \hat q_B+\hat H_G^{\rm local}.
\end{align}
The projected nonlocal coupling vanishes at the magic angle $3\cos^2\theta=1$. This angle is therefore a null of the gravitational coupling between the selected mechanical modes. At $\theta_m$, only the direct coupling between the selected modes vanishes; the mixed couplings to orthogonal modes remain finite. The null therefore requires those modes to be sufficiently detuned or independently characterized so that any residual effective coupling induced in the selected-mode dynamics lies below the calibration uncertainty.

\section{Cavity elimination}
\label{apdx:adiabatic}

We first determine the displacement dependence of the common cavity resonance. Let $\ell$ be the total one-way equilibrium optical path length between the two fixed end mirrors. At a fixed angle $\theta$, the common normal to the moving mirrors and the tangent in the $x$-$y$ plane are $\bm e_q = (\sin\theta,\cos\theta)^{\rm T}$ and $\bm e_\perp = (\cos\theta,-\sin\theta)^{\rm T}$. We take the propagation direction from mirror $A$ to mirror $B$ to be $-\bm e_y$. Specular reflection then fixes the common direction of the incoming and outgoing outer segments as $\bm s = (\sin2\theta,\cos2\theta)^{\rm T}$ and $\bm s + \bm e_y = 2\cos\theta \bm e_q$. For small mirror displacements $\bm q_j = \hat q_j\bm e_q+\hat q_j^\perp\bm e_\perp$, the first order path-length change is
\begin{align}
    \delta \ell &= 
    \bm s \cdot \bm q_A
    + \bm e_y \cdot (\bm q_A-\bm q_B)-\bm s \cdot \bm q_B\nonumber\\
    &= 2(\hat q_A-\hat q_B)\cos\theta.
\end{align}
Hence, displacements tangent to ideal plane mirrors make no contribution. At $\theta =0$, the folded path retraces itself. For the longitudinal example, we instead use the moving masses as the end mirrors of a Fabry-P\'erot cavity. Its one-way equilibrium length is $d$, its length change is $\hat q_A-\hat q_B$, and its linearized coupling is $\mathscr G(0) = \bar a \omega_{\rm cav}/d$. The elimination below applies to either configuration using its corresponding coupling.

Let $\hat a$ denote the cavity annihilation operator before linearization. The optomechanical Hamiltonian of the cavity mode is
\begin{align}
    \hat H_{\rm cav}=\hbar \omega_{\rm eff}(\hat{\bm r}_A,\hat{\bm r}_B)\hat a^\dagger \hat a,
\end{align}
where $\omega_{\rm eff}$ differs from the cavity resonance frequency $\omega_{\rm cav}$ because the effective cavity length changes due to radiation pressure. For the selected one-dimensional modes $\hat q_j$, we have
\begin{align}
    \omega_{\rm eff}(\hat{\bm r}_A,\hat{\bm r}_B)=\frac{\omega_{\rm cav}}{1+2(\hat q_A-\hat q_B)\cos\theta/\ell}.
\end{align}
After displacing the optical field, shifting the static mechanical equilibrium, and retaining the bilinear fluctuation terms relevant to the linearized dynamics, the Hamiltonian becomes
\begin{align}
    \hat H_{\rm cav}\simeq \hbar \omega_{\rm cav} \hat a^\dagger\hat a-\hbar \mathscr G(\theta)(\hat q_A-\hat q_B)(\hat a+\hat a^\dagger).
\end{align}
Here, $\bar a$ is chosen real, $\bar a^2$ is the intracavity photon number, and $\hat a$ denotes the displaced fluctuation operator. The first term is the harmonic part of the cavity mode, while the second is the optomechanical interaction with coupling $\mathscr G(\theta)=2\omega_{\rm cav}\bar a\cos\theta/\ell$; omitted local mechanical terms are included in the dressed mechanical Hamiltonian.

In the rotating frame at a laser frequency $\omega_L$, the equation of motion of the cavity mode is
\begin{align}
    \dot{\hat a}&=i\Delta \hat a+i\mathscr G(\theta)(\hat q_A-\hat q_B)-\kappa \hat a+\sqrt{2\kappa}\hat a_{\rm in},
\end{align}
where $\Delta=\omega_L-\omega_{\rm cav}$ is the cavity detuning, $\kappa$ is the decay rate of the cavity mode, and $\hat a_{\rm in}$ is the input optical field. Here, we assume the adiabatic limit, in which $\kappa$ is greater than the other dynamical rates. Then, we have
\begin{align}
    \hat a=\frac{1}{\kappa-i\Delta}(i\mathscr G(\theta)(\hat q_A-\hat q_B)+\sqrt{2\kappa}\hat a_{\rm in}).
\end{align}
Substituting this into the radiation pressure force $\hbar \mathscr G(\theta)(\hat a+\hat a^\dagger)$ in the mechanical motion gives
\begin{align}
    &-\frac{2\hbar \Delta \mathscr G^2(\theta)}{\kappa^2+\Delta^2}(\hat q_A-\hat q_B)\nonumber\\
    &\quad+\hbar \mathscr G(\theta)\left(\frac{\sqrt{2\kappa}}{\kappa-i\Delta}\hat a_{\rm in}+\frac{\sqrt{2\kappa}}{\kappa+i\Delta}\hat a_{\rm in}^\dagger\right),
\end{align}
where the first line is the coherent caivity-mediated force, while the second line is the correlated optical backaction force. The local pieces of the coherent term renormalize the mechanical frequencies and are absorbed into the dressed local Hamiltonian, whereas its cross term generates the effective nonlocal coupling $K_E\hat q_A\hat q_B$. The resulting additional coupling and backaction spectra are
\begin{align}
    K_E(\theta)&=-\frac{2\hbar \Delta \mathscr G^2(\theta)}{\kappa^2+\Delta^2}\\
    S_{F_A^EF_A^E}(\theta)&=S_{F_B^EF_B^E}(\theta)=-S_{F_A^EF_B^E}(\theta)=\frac{4\hbar^2\mathscr G^2(\theta)\kappa}{\kappa^2+\Delta^2}S_{a_{\rm in}a_{\rm in}},
\end{align}
where $S_{a_{\rm in}a_{\rm in}}$ is the spectral density of the input optical field. In particular, for vacuum input, we have $S_{a_{\rm in}a_{\rm in}}=1/2$.

\section{Pulsed protocol}
\label{apdx:pulsed protocol}

We describe a pulsed realization of preparation and readout and derive the output entanglement, following Ref.~\cite{miki2026}. The first pulses transfer optical states to the mechanical modes. The two masses then evolve under gravity and the additional interaction for a duration time $t_G$ in the presence of force noise, and the second pulses transfer the final mechanical states onto outgoing optical modes. The first stage is an illustrative preparation method: the covariance evolution also applies to locally prepared conditional mechanical states.

We consider identical mechanical masses with mass $M$ and frequency $\omega_m$, with phonon operators $\hat b=\sqrt{M\omega_m/2\hbar}\hat q_j+i/\sqrt{2\hbar M\omega_m}\hat p_j$. In the interaction picture, the RWA gives the local optomechanical interaction Hamiltonian
\begin{align}
    \hat H_{\rm opt-int,loc}^I&=
    -\hbar \sum_{j=A,B}g_j(t)(\hat c_j^\dagger\hat b_j+\hat c_j\hat b_j^\dagger),
\end{align}
where $\hat c_j$ is the local cavity annihilation operator, $[\hat c_j,\hat c_k^\dagger]=\delta_{jk}$, and $g_j(t)$ is the local optomechanical coupling strength. We take the local cavities to be red detuned, $\Delta_{\rm loc}=-\omega_m$, and assume the resolved-sideband and adiabatic-cavity hierarchy, $\omega_m\gg\kappa_{j,\rm loc}\gg g_j$. Here $\kappa_{j,\rm loc}$ is the cavity decay rate. Both pulsed stages obey
\begin{align}
    \dot{\hat b}_j&=ig_j \hat c_j,\quad
    \dot{\hat c}_j=ig_j\hat b_j-\kappa_{j,\rm loc}\hat c_j+\sqrt{2\kappa_{j,\rm loc}}\hat c_{j,\rm in}\\
    \hat c_{j,\rm out}&=\sqrt{2\kappa_{j,\rm loc}}\hat c_j-\hat c_{j,\rm in},
\end{align}
where $\hat c_{j,\rm in}$ and $\hat c_{j,\rm out}$ are the input and output optical fields satisfying $[\hat c_{j,\rm in}(t),\hat c_{j,\rm in}^\dagger(t^\prime)]=[\hat c_{j,\rm out}(t),\hat c_{j,\rm out}^\dagger(t^\prime)]=\delta(t-t^\prime)$. We take $g_j(t)=g_j$ during $0\le t<t_P$ and $t_P+t_G\le t <2t_P+t_G$, and zero during the intervening mechanical interaction stage. We assume that gravity, the additional intermass interaction, mechanical damping, and force noise have negligible effects during each short pulse.
We define the normalized temporal mode of the first pulse as
\begin{align}
    \hat \xi_{j,\rm in}^{(1)}&=
    \sqrt{\frac{2G_j}{e^{2G_jt_P}-1}}\int_0^{t_P}ds\,e^{G_js}\hat c_{j,\rm in}(s)\\
    \hat \xi_{j,\rm out}^{(1)}&=
    \sqrt{\frac{2G_j}{1-e^{-2G_jt_P}}}\int_0^{t_P}ds\,e^{-G_js}\hat c_{j,\rm out}(s),
\end{align}
and the second pulse as
\begin{align}
    \hat \xi_{j,\rm in}^{(2)}&=
    \sqrt{\frac{2G_j}{e^{2G_jt_P}-1}}\int_{t_P+t_G}^{2t_P+t_G}ds\,e^{G_j(s-t_P-t_G)}\hat c_{j,\rm in}(s)\\
    \hat \xi_{j,\rm out}^{(2)}&=
    \sqrt{\frac{2G_j}{1-e^{-2G_jt_P}}}\int_{t_P+t_G}^{2t_P+t_G}ds\,e^{-G_j(s-t_P-t_G)}\hat c_{j,\rm out}(s),
\end{align}
where $[\hat \xi_{j,\rm in}^{(1)/(2)},\hat \xi_{j,\rm in}^{(1)/(2)\dagger}]=[\hat \xi_{j,\rm out}^{(1)/(2)},\hat \xi_{j,\rm out}^{(1)/(2)\dagger}]=1$, and we define $G_j=g_j^2/\kappa_{j,\rm loc}$. For $G_jt_P\gg1$, transfer approaches a complete swap. We use this ideal limit: the first pulses prepare the mechanical modes in the input optical states, and the second pulses transfer the final mechanical states to the output optical modes.
The covariance matrices are therefore
\begin{align}
    \bm V_b(t_P)=\bm V_{\xi^{\rm in}},\quad
    \bm V_{\xi^{\rm out}}=\bm V_b(t_P+t_G).
\end{align}

Next, we consider the gravitational and additional interaction stage in the presence of thermal noise and backaction. The Hamiltonian in the interaction picture is
\begin{align}
    \hat H_I(\theta)=
    \hbar g_\theta
    \left(
    \hat b_A\hat b_B^\dagger+
    \hat b_A^\dagger\hat b_B
    \right),\qquad
    g_\theta
    =
    \frac{K_E+K_G(\theta)}{2M\omega_m}.
    \label{eq:app RWA Hamiltonian}
\end{align}
Here the RWA is taken for $|g_\theta|\ll\omega_m$, neglecting the counter-rotating terms that oscillate at $2\omega_m$.
The equations of motion are
\begin{align}
    \dot{\hat b}_A=-ig_\theta\hat b_B-\frac{\gamma}{2}\hat b_A+\mathscr F_{A,\rm nor}\\
    \dot{\hat b}_B=-ig_\theta\hat b_A-\frac{\gamma}{2}\hat b_B+\mathscr F_{B,\rm nor},
\end{align}
where $\mathscr F_{j,\rm nor}=ie^{i\omega_mt}\mathscr F_j/\sqrt{2M\hbar \omega_m}$ and $\langle\{\mathscr F_j(t),\mathscr F_k(t^\prime)\}\rangle/2=S_{\mathscr F_j\mathscr F_k}\delta(t-t^\prime)$ with
\begin{align}
    S_{\mathscr F_A\mathscr F_A}&=
    2M\gamma k_BT+S_{F_A^EF_A^E}\\
    S_{\mathscr F_B\mathscr F_B}&=
    2M\gamma k_BT+S_{F_B^EF_B^E}\\
    S_{\mathscr F_A\mathscr F_B}&=
    S_{F_A^EF_B^E}
\end{align}
with $N_{\rm th}=k_BT/\hbar \omega_m$ in the high-temperature limit. The mechanical linewidth $\gamma$ is defined as the full width at half maximum, so that $\gamma=2\gamma_m$ in the convention of Ref.~\cite{miki2026}. We define the quadratures $\hat b_{1j}=\hat b_j+\hat b_j^\dagger$ and $\hat b_{2j}=(\hat b_j-\hat b_j^\dagger)/i$. The covariance matrix $\bm V_b=\langle\{\hat{\bm r}-\langle\hat{\bm r}\rangle,(\hat{\bm r}-\langle\hat{\bm r}\rangle)^{\rm T}\}\rangle/2$, with $\hat{\bm r}=(\hat b_{1A},\hat b_{2A},\hat b_{1B},\hat b_{2B})^{\rm T}$, obeys
\begin{align}
    \bm V_b(t_P+t_G)=
    e^{-\gamma t_G}
    \bm{R}_\theta(t_G)
    \bm V_b(t_P)
    \bm{R}_\theta^T(t_G)+
    \bm V_N,
    \label{eq:apdx output covariance}
\end{align}
where $\bm V_b(t_P)$ is the covariance matrix at the start of the interaction stage whether prepared by optical state transfer. For conditional preparation, it is the covariance conditioned on the retained measurement records.
The rotation $\bm{R}_\theta$ and the accumulated noise $\bm V_N$ are
\begin{align}
    \bm{R}_\theta(t)&=
    \left(\begin{array}{cccc}
    \cos[g_\theta t] & 0 & 0 & \sin[g_\theta t] \\
    0 & \cos[g_\theta t] & -\sin[g_\theta t] & 0 \\
    0 & \sin[g_\theta t] & \cos[g_\theta t] & 0 \\
    -\sin[g_\theta t] & 0 & 0 & \cos[g_\theta t]
    \end{array}\right)\\
    \bm V_N&=
    \int_0^{t_G}
    d\tau\,
    e^{-\gamma\tau}
    \bm{R}_\theta(\tau)
    \bm D
    \bm{R}_\theta^T(\tau)\\
    &=I_0
    \begin{pmatrix}
    \frac{S_{\mathscr F_A\mathscr F_A}}
    {M\hbar\omega_m} & 0 & \frac{S_{\mathscr F_A\mathscr F_B}}{M\hbar\omega_m} & 0 \\
    0 & \frac{S_{\mathscr F_A\mathscr F_A}}{M\hbar\omega_m} & 0 & \frac{S_{\mathscr F_A\mathscr F_B}}{M\hbar\omega_m} \\
    \frac{S_{\mathscr F_A\mathscr F_B}}{M\hbar\omega_m} & 0 & \frac{S_{\mathscr F_A\mathscr F_A}}{M\hbar\omega_m} & 0 \\
    0 & \frac{S_{\mathscr F_A\mathscr F_B}}{M\hbar\omega_m} & 0 & \frac{S_{\mathscr F_A\mathscr F_A}}{M\hbar\omega_m}
    \end{pmatrix},
\end{align}
where $\bm D$ is the quadrature diffusion matrix, $I_0=(1-e^{-\gamma t_G})/\gamma$, and for identical systems with equal auto-noise spectra, $S_{\mathscr F_A\mathscr F_A}=S_{\mathscr F_B\mathscr F_B}$. In this case $\bm R_\theta \bm D \bm R_\theta^{\rm T} = \bm D$, which gives the last equality.

We introduce the entanglement negativity for the two-mode Gaussian states as
\begin{align}
    \mathscr N_G={\rm max}\{(1/\nu_--1)/2,0\},
\end{align}
where $\nu_-$ is the smallest symplectic eigenvalue after partial transpose, given by $\nu_-^2=(\sigma-\sqrt{\sigma^2-4\det\bm V_b(t_P+t_G)})/2$ with $\sigma=\det\bm V_{b,A}+\det\bm V_{b,B}-2\det\bm V_{b,AB}$. Here $\bm V_{b,A}$ and $\bm V_{b,B}$ are the local covariance blocks, and $\bm V_{b,AB}$ is the correlation block of $\bm V_b$. The state is entangled if and only if $\mathscr N_G>0$.

For the analytic example, we assume an initial product of two pure mechanical squeezed states, $\bm V_{b,\rm in}={\rm diag}\{e^{2\zeta},e^{-2\zeta},e^{2\zeta},e^{-2\zeta}\}$. We choose this family because, when the interaction time is selected to maximize the negativity, its entanglement threshold approaches the universal separability-preservation bound in the limit of infinite squeezing. Violation of the universal bound does not guarantee entanglement for a given initial state; other initially separable states may fail to generate it altogether. For example, a product of vacuum states remains separable under the RWA evolution with independent thermal noise even when $N_{\mathscr F}<\hbar|K_G+K_E|$.

In the regime $\gamma t_G\ll1$, we obtain
\begin{align}
    \nu_-^2&\simeq
    \left(\cosh[2\zeta]+\frac{t_GS_{\mathscr F_A\mathscr F_A}}{M\hbar\omega_m}\right)^2-
    \cos^2[2g_\theta t_G]\sinh^2[2\zeta]\nonumber\\
    &\quad
    +\sinh^2[2\zeta]\sin^2[2g_\theta t_G]
    -\left(\frac{t_GS_{\mathscr F_A\mathscr F_B}}{M\hbar\omega_m}\right)^2\nonumber\\
    &\quad
    -\biggl\{4\sinh^2[2\zeta]\cos^2[2g_\theta t_G]\left(\frac{t_GS_{\mathscr F_A\mathscr F_B}}{M\hbar\omega_m}\right)^2\nonumber\\
    &\quad
    +4\left(\left(\cosh[2\zeta]+\frac{t_GS_{\mathscr F_A\mathscr F_A}}{M\hbar\omega_m}\right)^2-\cos^2[2g_\theta t_G]\sinh^2[2\zeta]\right)\nonumber\\
    &\quad\times
    \sinh^2[2\zeta]\sin^2[2g_\theta t_G]\biggr\}^{1/2}
\end{align}
When the thermal noise is significantly larger than the backaction, we have
\begin{align}
    \nu_-^2&\simeq
    \left(\cosh[2\zeta]+2\gamma N_{\rm th}t_G\right)^2-
    \cos^2[2g_\theta t_G]\sinh^2[2\zeta]\nonumber\\
    &\quad
    +\sinh^2[2\zeta]\sin^2[2g_\theta t_G]\nonumber\\
    &\quad
    -\biggl\{
    4\left(\left(\cosh[2\zeta]+2\gamma N_{\rm th}t_G\right)^2-\cos^2[2g_\theta t_G]\sinh^2[2\zeta]\right)\nonumber\\
    &\quad\times
    \sinh^2[2\zeta]\sin^2[2g_\theta t_G]\biggr\}^{1/2}\nonumber\\
    &=\biggl\{\sqrt{\left(\cosh[2\zeta]+2\gamma N_{\rm th}t_G\right)^2-
    \cos^2[2g_\theta t_G]\sinh^2[2\zeta]}\nonumber\\
    &\quad
    -|\sinh[2\zeta]\sin[2g_\theta t_G]|\biggr\}^2,
\end{align}
which reproduces the result of Ref.~\cite{miki2026} upon replacing $g_G$ with $g_\theta$.

\begin{figure}
    \centering
    \includegraphics[width=0.95\linewidth]{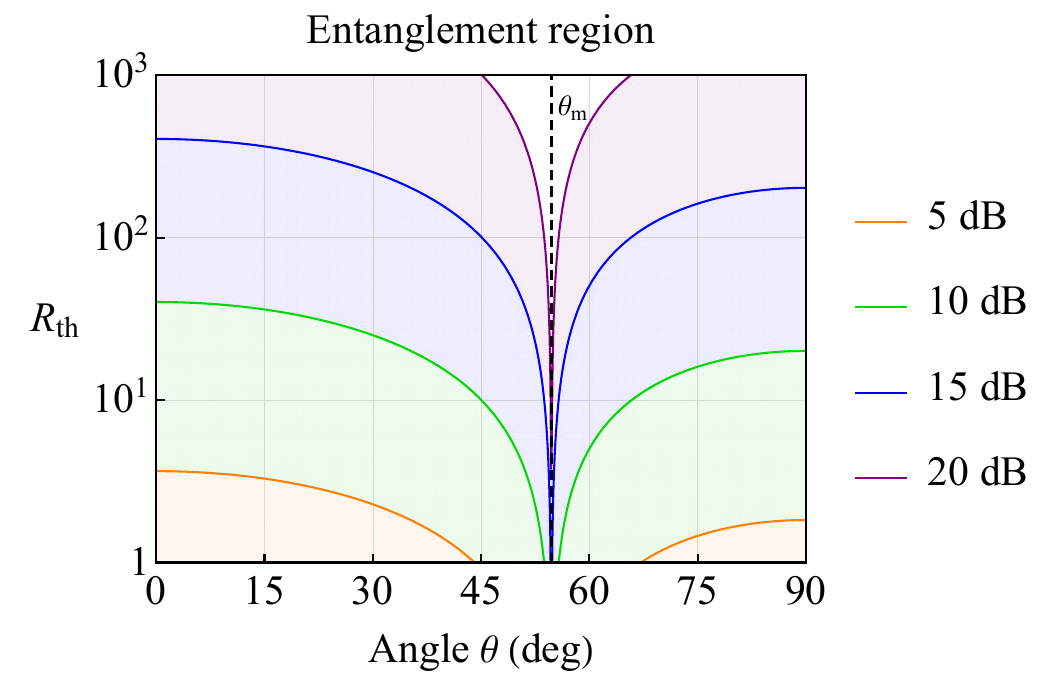}
    \caption{
    Entangling regions for pure initial mechanical squeezed states, with $K_E = K_E^{\rm mid}$, $M/d^3=20~{\rm g}/{\rm cm}^3$, and thermal noise dominating optical background. The black vertical dashed line marks the magic angle $\theta_m$. Entanglement occurs below each corresponding curve, including the lower colored bands.
    }
    \label{fig:entanglement region}
\end{figure}

\section{Angular dependence and optimal interaction time}
\label{apdx:angular dependence}

We compare the entanglement regions for different angles and initial mechanical squeezing, and explain the time optimization used in Fig.~\ref{fig:en}. At each non-null angle, we set $K_E=K_E^{\rm mid}$ and require a magic-angle reference with matching additional coupling and force-noise spectra. This coupling is retuned between operating points and held fixed during the time optimization at each point.

Figure~\ref{fig:entanglement region} shows the entangling regions for $5$, $10$, $15$, and $20~{\rm dB}$ of initial mechanical squeezing. Each solid curve satisfies $\tanh2\zeta = \gamma N_{\rm th}/|g_\theta|$, the entanglement threshold obtained from Eq.~\eqref{eq:symplectic eigenvalue} after optimizing the time. Increasing the squeezing enlarges the entangling region. The noise tolerance is largest at $\theta = 0$ and decreases towards the magic angle.

The entanglement negativity determined by Eq.~\eqref{eq:symplectic eigenvalue} depends on the three quantities $\zeta$, $g_\theta t_G$, and $\gamma t_GN_{\rm th}$. Substituting $K_E = K_E^{\rm mid}$, $K_G(\theta)/K_0 = 3\cos^2\theta-1$, $N_{\mathscr F}\simeq 2M\gamma k_BT$, and $R_{\rm th} = \gamma T/\gamma_0T_0$, the last two terms are
\begin{align}
    g_\theta t_G &= {\rm sgn}[3\cos^2\theta-1] \nonumber \\
    &\quad\times
    \left[\frac{(GM/d^3)|3\cos^2\theta-1|}{4}+\frac{k_B\gamma_0T_0R_{\rm th}}{\hbar}\right]\frac{t_G}{\omega_m} \nonumber \\
    \gamma t_GN_{\rm th} &= \frac{k_B\gamma_0T_0R_{\rm th}}{\hbar} \frac{t_G}{\omega_m}
    \label{eq:coherent phase}
\end{align}
At each fixed $\zeta$, $\theta$, $R_{\rm th}$, and $M/d^3$, the remaining variable is $t_G/\omega_m$. We choose it to maximize $\mathscr N_G$, varying the coherent phase and the accumulated thermal noise together according to Eq.~\eqref{eq:coherent phase}.
It is sufficient to search $0 \le |g_\theta t_G| \le \pi/4$: every later phase has the same coherent trigonometric factors as an earlier point in this interval, but accumulates more thermal noise. If the minimum value of $\nu_-$ is not below unity, the maximum negativity is zero.

\end{document}